\documentclass{LMCS}

\usepackage[latin1]{inputenc}
\usepackage{amssymb}
\usepackage{graphicx,color}
\usepackage{enumerate}
\usepackage{hyperref}

\newcommand{\reals}{\mathbb{R}}
\newcommand{\ints}{\mathbb{Z}}
\newcommand{\rationals}{\mathbb{Q}}
\newcommand{\nats}{\mathbb{N}}

\newcommand{\lcm}{\,\mbox{lcm}\,}
\newcommand{\cal}{\mathcal}

\newtheorem{property}[thm]{Property}

\def\doi{6 (1:6) 2010}
\lmcsheading%
{\doi}
{1--17}
{}
{}
{Dec.~12, 2008}
{Feb.~24, 2010}
{}   

\begin{document}

\title[On the Sets of Real Numbers Recognized by Finite Automata]
{On the Sets of Real Numbers Recognized by Finite Automata in
  Multiple Bases\rsuper*}

\author[B. Boigelot]{Bernard Boigelot\rsuper a}
\address{{\lsuper{a,b}}Universit\'e de Li\`ege\\
Institut Montefiore, B28\\
B-4000 Li\`ege, Belgium
}
\email{\{boigelot,brusten\}@montefiore.ulg.ac.be}

\author[J.~Brusten]{Julien Brusten\rsuper b}
\vskip-6 pt
\thanks{{\lsuper b}Research fellow (``Aspirant'') of the Belgian Fund for Scientific 
Research (F.R.S.-FNRS)}

\author[V.~Bruy\`ere]{V\'eronique Bruy\`ere\rsuper c}
\address{{\lsuper c}Universit\'e de Mons-Hainaut\\
Avenue du Champ de Mars, 6\\
B-7000 Mons, Belgium
}
\email{veronique.bruyere@umh.ac.be}

\keywords{automata, real numbers, mixed real-integer arithmetic, Cobham's
theorem}
\subjclass{F.1.1; F.4.1; F.4.3.}
\titlecomment{{\lsuper*}A preliminary version of this work appears in the proceedings
of the 35th International Colloquium on Automata, Languages and
Programming (ICALP'08).  This work is supported by
the {\em Interuniversity Attraction Poles\/} program {\em MoVES\/} of
the Belgian Federal Science Policy Office, and by the grant 2.4530.02
of the Belgian Fund for Scientific Research (F.R.S.-FNRS)}

\begin{abstract}
This article studies the expressive power of finite automata recognizing
sets of real numbers encoded in positional notation. We consider
Muller automata as well as the restricted class of {\em
  weak deterministic} automata, used as symbolic set representations
in actual applications. In previous work, it has been established that
the sets of numbers that are recognizable by weak deterministic
automata in two bases that do not share the same set of prime factors
are exactly those that are definable in the first order additive
theory of real and integer numbers.
This result extends {\em Cobham's theorem\/}, which
characterizes the sets of integer numbers that are recognizable by
finite automata in multiple bases.

In this article, we first generalize this result to {\em
  multiplicatively independent\/} bases, which brings it closer to the
original statement of Cobham's theorem. Then, we study the sets of
reals recognizable by Muller automata in two bases. We show with a
counterexample that, in this setting, Cobham's theorem does not
generalize to multiplicatively independent bases. Finally, we prove
that the sets of reals that are recognizable by Muller automata
in two bases that do not share the same set of prime factors are
exactly those definable in the first order additive theory of
real and integer numbers. 
These sets are thus also recognizable by weak deterministic
automata.  This result leads to a precise characterization of the sets
of real numbers that are recognizable in multiple bases, and provides
a theoretical justification to the use of weak automata as symbolic
representations of sets.
\end{abstract}

\maketitle

\section{Introduction}
By using positional notation, real numbers can be encoded as
infinite words over an alphabet composed of a fixed number of digits,
with an additional symbol for separating their integer and fractional
parts. This encoding scheme maps sets of numbers onto languages that
describe those sets unambiguously.

This article studies the sets of real numbers whose encodings can be
accepted by finite automata. The motivation is twofold.  First, since
regular languages enjoy good closure properties under a large range of
operators, automata provide powerful theoretical tools for
establishing the decidability of arithmetic theories. In particular,
it is known that the sets of numbers that are definable in the
first-order additive theory of integers $\langle \ints, +, < \rangle$,
also called {\em Presburger arithmetic\/}, are encoded by regular
finite-word languages~\cite{Buc62,BHMV94}. This result translates into
a simple procedure for deciding the satisfiability of Presburger
formulas. Moving to infinite-word encodings and $\omega$-regular
languages, it can be extended to sets of real numbers definable in
$\langle \reals, \ints, +, < \rangle$, i.e., the first-order additive
theory of real and integer variables~\cite{BBR97,BRW98,BJW05}.

The second motivation is practical. Since finite automata are objects
that are easily manipulated algorithmically, they can be used as
actual data structures for representing symbolically sets of
values. This idea has successfully been exploited in the context of
computer-aided verification, leading to representations suited for the
sets of real and integer vectors handled during symbolic state-space
exploration~\cite{WB95,Boi98,BJW05,EK06}. A practical limitation of this
approach is the high computational cost of some operations involving
infinite-word automata, in particular language
complementation~\cite{Saf88,Var07}.  However, it has been shown that a
restricted form of automata, {\em weak deterministic\/} ones, actually
suffices for handling the sets definable in $\langle \reals, \ints, +,
< \rangle$~\cite{BJW05}. Weak automata can be manipulated with
essentially the same cost as finite-word ones~\cite{Wil93}, which
alleviates the problem and leads to an effective representation
system.

Whether a set of numbers can be recognized by an automaton generally
depends on the chosen encoding base. For integer numbers, it is known
that a set $S \subseteq \ints$ is recognizable in a base $r > 1$ iff
it is definable in the theory $\langle \ints, +, {<,} V_r \rangle$,
where $V_r$ is a base-dependent function~\cite{BHMV94} that returns
the highest integer power of $r$ dividing its argument. Furthermore,
the well-known {\em Cobham's theorem\/} states that if a set $S
\subseteq \nats$ is simultaneously recognizable in two bases $r > 1$
and $s > 1$ that are {\em multiplicatively independent\/}, i.e., such
that $r^p \neq s^q$ for all $p, q \in \nats_{>0}$, then $S$ is {\em
  ultimately periodic\/}, i.e., it differs from a periodic subset of
$\nats$ only by a finite set~\cite{Cob69}. As a corollary of Cobham's
theorem, a subset  of $\ints$ that is recognizable in two
multiplicatively independent bases is definable in $\langle \ints, +,
< \rangle$~\cite{BHMV94}, from which it follows that it is
recognizable in every base.  Our aim is to generalize as completely as
possible this result to automata recognizing real numbers, by
precisely characterizing the sets that are recognizable in multiple
bases. We first consider the case, relevant for practical
applications, of weak deterministic automata. In previous work, it has
been established that a set of real numbers is simultaneously
recognizable by weak deterministic automata in two bases that do not
share the same set of prime factors iff this set is definable in
$\langle \reals, \ints, +, < \rangle$~\cite{BB09}.  As a first
contribution, we extend this result to pairs of multiplicatively
independent bases. Since recognizability in two multiplicatively
dependent bases is equivalent to recognizability in only one of
them~\cite{BRW98}, this result provides a complete characterization of
the sets that are recognizable in multiple bases by weak deterministic
automata.

Then, we move to sets recognized by Muller
automata. We establish that there exists a set of real numbers
recognizable in two multiplicatively independent bases that share the
same set of prime factors, but that is not definable in $\langle
\reals, \ints, +, < \rangle$. This shows that Cobham's theorem does
not directly generalize to Muller automata
recognizing sets of real numbers. Finally, we establish that a set $S
\subseteq \reals$ is simultaneously recognizable in two bases that do
not share the same set of prime factors iff $S$ is definable in
$\langle \reals, \ints, +, < \rangle$. As a corollary, such a set must
then be recognizable by a weak deterministic automaton. Our result
thus provides a theoretical justification to the use of weak automata,
by showing that their expressive power corresponds precisely to the
sets of reals recognizable by infinite-word automata in every
encoding base.

\section{Basic notions}

\subsection{Encodings of numbers}

Let $r \in \nats_{>1}$ be an integer numeration {\em base\/} and let $\Sigma_r =
\{0, \ldots, r-1\}$ be the corresponding set of {\em digits\/}.  We
encode a real number $x$ in base $r$, most significant digit first, by
words of the form $w_I \star w_F$, where $w_I \in
\Sigma_r^*$ encodes an integer part $x_I \in \ints$ of $x$ and $w_F \in
\Sigma_r^\omega$ encodes a fractional part $x_F \in [0,1]$.  Note that
the decomposition of $x$ into $x_I$ and $x_F$ is not necessarily unique, e.g.,
$x = 3$ gives either $x_I = 3$ and $x_F = 0$, or $x_I=2$ and $x_F=1$.
Negative integer numbers are represented by their $r$'s-complement, i.e., the
encodings of $x_I \in \ints_{<0}$ are formed by the last $p$ 
digits of the encodings of $r^p + x_I$.  The length $p$ of $w_I$ is not
fixed but has to be large enough for $-r^{p-1} \leq x_I < r^{p-1}$ to
hold; thus, the most significant digit of an encoding 
is equal to $0$ for positive integer parts and to $r-1$ for negative
ones~\cite{BBR97}. As a consequence, the set of {\em valid encodings\/}
of numbers in base $r$ forms the language $\{ 0, r-1 \} \Sigma_r^*
\star \Sigma_r^{\omega}$.
  Some numbers have two distinct encodings with the
same integer-part length, e.g., in base $10$, the number $11/2$ admits
the encodings $0^+ 5 \star 5 0^\omega$ and $0^+ 5 \star 4
9^\omega$. Such encodings are called {\em dual\/}.
For a word $w = b_{p-1}^I b_{p-2}^I \ldots b_1^I b_0^I
\star b_1^F b_2^F b_3^F \ldots \in \{0,r-1\} \Sigma_r^* \star
\Sigma_r^\omega$, we denote by $[w]_r$ the real number encoded by $w$
in base $r$, i.e.,
\[[w]_r = \sum_{i=0}^{p-2} b_i^I r^i + \sum_{i > 0} b_i^F r^{-i} + 
\left\{
\begin{array}{cl}
0 & \mbox{ if } b_{p-1}^I = 0,\\
-r^{p-1} & \mbox{ if } b_{p-1}^I = r - 1.
\end{array}
\right.
\]
For finite
words $w \in \Sigma_r^*$, we denote by $[w]_r$ the natural number
encoded by $w$, i.e., $[w]_r = [0 w \star 0^{\omega}]_r$.

It is known~\cite{HW85} that a word $w \in \{0,r-1\}\Sigma_r^*\star
\Sigma_r^\omega$ is ultimately periodic, i.e., of the form
$\{0,r-1\}u_1\star u_2u_3^\omega$ with $u_1, u_2 \in \Sigma_r^*$ and
$u_3 \in \Sigma_r^+$, if and only if $[w]_r$ is rational. The word
$u_3$ is then called a {\em period\/} of $w$.

\subsection{Real Number Automata}
For a set $S \subseteq \reals$, we denote by $L_r(S)$ the language of
all the base-$r$ encodings of the elements of $S$.
If $L_r(S)$ is $\omega$-regular, then it can be
accepted by a (non-unique) infinite-word automaton, called a {\em Real
Number Automaton (RNA)\/}, recognizing $S$. Such a set $S$ is then
said to be {\em $r$-recognizable\/}.  RNA can be generalized into {\em
Real Vector Automata (RVA)\/}, suited for subsets of $\reals^n$,
with $n > 0$~\cite{BBR97}.

RNA have originally been defined as B\"uchi
automata~\cite{BBR97}. In this article, we will instead consider them to
be {\em deterministic Muller\/} automata. This adaptation can be made 
without loss of generality, since both classes of automata share the same
expressive power~\cite{McN66,PP04}. The fact that RNA have a
deterministic transition relation will simplify technical
developments.

The $r$-recognizable sets of real numbers are precisely described
by the following result. This logical characterization
will often be used in this article.

\begin{thm}[\cite{BRW98}]
\label{theo-brw98}
Let $r \in \nats_{>1}$ be a base.  A subset of $\reals$ is
$r$-recognizable iff it is definable in the first-order theory
$\langle \reals, \ints, +, <, X_r \rangle$, where $X_r(x,u,k)$ is a
base-dependent predicate that holds whenever $u$ is an integer power
of $r$, and there exists an encoding of $x$ in which the digit at the
position specified by $u$ is equal to $k$.
\end{thm}

It is known that the full expressive power of infinite-word automata
is not needed for representing the subsets of $\reals$ that are
definable in $\langle \reals, \ints, +, < \rangle$. The following
theorem establishes that such sets can be recognized by {\em weak
  deterministic\/} automata, i.e., deterministic B\"uchi automata such
that each strongly connected component of their transition graph
contains either only accepting or only non-accepting
states.  A set recognized by a weak deterministic
automaton in base $r$ is said to be {\em weakly $r$-recognizable\/},
and such an automaton is then called a {\em weak RNA\/}.

\begin{thm}[\cite{BJW05}]
\label{theo-bjw05}
If a subset of $\reals$ is definable in the first-order theory
$\langle \reals, \ints, +, < \rangle$, then it is weakly $r$-recognizable in
every base $r \in \nats_{>1}$.
\end{thm}

\subsection{Topology}

In this section, we recall some notions about topology, which is a
useful tool for reasoning about the properties of sets of words and 
numbers~\cite{PP04}.

\subsubsection{General concepts}

Given a set $S$, either of words or of numbers, 
a distance $d(x,y)$ defined on
this set induces a metric topology on subsets of $S$.  A 
{\em neighborhood\/}
$N_\varepsilon(x)$ of a point $x \in S$ with respect to $\varepsilon
\in \reals_{> 0}$ is the set $N_{\varepsilon}(x) = 
\{y \mid d(x,y) < \varepsilon\}$.  A set $C \subseteq S$ is said
to be {\em open\/} if for all $x \in C$, there exists $\varepsilon > 0$
such that $N_\varepsilon(x) \subseteq C$.  A {\em closed\/} set is
a set whose complement with respect to $S$ is open, or, equivalently,
a set that contains the limits of all its converging sequences 
of elements.  The following notations will be used:

\begin{enumerate}[$\bullet$]
\item $F$ is the class of closed sets,
\item $G$ is the class of open sets,
\item $F_{\sigma}$ is the class of countable unions of closed sets,
\item $G_{\delta}$ is the class of countable intersections of open
sets.
\end{enumerate}

Other classes can be defined from these notations: The class
${\cal B}(F) = {\cal B}(G)$ contains the finite Boolean combinations
of open and closed sets, whereas $F_\sigma \cap G_\delta$ is the class
of sets that can be expressed as countable unions of closed sets
as well as countable intersections of open sets.

Those classes of sets are the first levels of the Borel hierarchy.  
In a metric topology, this hierarchy states that $F$ and $G$ are 
subclasses of ${\cal B}(F) = {\cal B}(G)$, which is itself
a subclass of $F_\sigma \cap G_\delta$.

\subsubsection{Topology of $\omega$-words}

Given a base $r \in \nats_{>1}$ and 
the alphabet $\Sigma_r \cup \{\star\}$, we define the following 
distance relation between infinite words over this alphabet:

\[ d(w,w') = \left\{ 
\begin{array}{cl}
\frac{1}{|common(w,w')|+1} & \mbox{ if } w \neq w'\\
0 & \mbox{ if } w = w',
\end{array}
\right.
\]
where $|common(w,w')|$ denotes the length of the longest common prefix of
$w$ and $w'$. This distance induces a topology on 
$(\Sigma_r \cup \{\star\})^\omega$.

We say that a $\omega$-language 
$L \subseteq (\Sigma_r \cup \{\star\})^\omega$ satisfies the 
{\em dense oscillating sequence property\/} if,
$w_1, w_2, w_3, \ldots$ being $\omega$-words and 
$\varepsilon_1, \varepsilon_2, \varepsilon_3, \ldots$ being distances, 
one has that
\[
\exists w_1 \forall \varepsilon_1 \exists w_2 \forall \varepsilon_2
\exists w_3 \forall \varepsilon_3 \cdots
\]
such that $d(w_i, w_{i+1}) \leq \varepsilon_i$ for all $i \geq 1$,
$w_i \in L$ for all odd $i$, and $w_i \notin L$ for all even 
$i$~\cite{BJW05}.

It has been established~\cite{MS97} that weak deterministic automata
accept exactly the $\omega$-regular languages that belong to the 
topological class $F_\sigma \cap G_\delta$.

It is also known~\cite{BJW05} that the $\omega$-regular languages
that satisfy the dense oscillating sequence property cannot be
accepted by weak deterministic automata.

\subsubsection{Topology of real numbers}

We consider the topology on the sets of real numbers induced by
the distance relation defined by $d(x,y) = |x - y|$.  

In this topology, a notion of dense oscillating sequence can
be defined in the same way as for $\omega$-words:
We say that a set $S \subseteq  \reals$
satisfies the {\em dense oscillating sequence property\/} if,
$x_1, x_2, x_3, \ldots$ being real numbers and 
$\varepsilon_1, \varepsilon_2, \varepsilon_3, \ldots$ being distances, 
one has that
\[
\exists x_1 \forall \varepsilon_1 \exists x_2 \forall \varepsilon_2
\exists x_3 \forall \varepsilon_3 \cdots
\]
such that $d(x_i, x_{i+1}) \leq \varepsilon_i$ for all $i \geq 1$,
$x_i \in S$ for all odd $i$, and $x_i \notin S$ for all even $i$.

We have the following theorem.

\begin{thm} \label{thm-dense-oscillating}
Let $r \in \nats_{>1}$ be a base.
The $r$-recognizable sets $S \subseteq \reals$ that satisfy the
dense oscillating sequence property are not weakly $r$-recognizable.
\end{thm}

\proof\
Consider a $r$-recognizable set $S \subseteq \reals$ satisfying
the dense oscillating sequence property. It is sufficient to establish that
$L_r(S)$ satisfies the dense oscillating sequence property as well.

Recall that each real number admits multiple encodings.  First, the first
digit of an encoding can be repeated at will.  Second, for a given length
of the integer part (assumed to be sufficiently large), a number admits
either one encoding, or two (dual) ones.

Let $S_1, S_2 \subseteq \reals$ be sets of numbers such that
$S_1 \cap S_2 = \emptyset$.  Consider any
number $x_1 \in S_1$ for which there exist arbitrarily close numbers
in $S_2$. Then, there exists an encoding $w_1$ of $x_1$ for which there
exist arbitrarily close encodings $w_2$ of numbers $x_2$ of $S$.  We
can ask more: There exists an encoding $w_1$ of $x_1$ for which
there exist arbitrarily close encodings $w_2$ of numbers $x_2$ of $S_2$,
including the dual encodings with the same integer part length as
$w_1$, if any.
Formally, $(x_1 \in S_1 \,\wedge\, (\forall
\varepsilon > 0)(\exists x_2 \in S_2) (d(x_1,x_2) < \varepsilon))
\,\Rightarrow\, (\exists w_1)([w_1]_r = x_1 \,\wedge\, (\forall
\varepsilon' > 0)(\exists x_2 \in S_2)( (\exists w_2)([w_2]_r = x_2
\,\wedge\, |w_1|_I = |w_2|_I) \,\wedge\, (\forall w_2)([w_2]_r = x_2
\,\wedge\, |w_1|_I = |w_2|_I \,\Rightarrow\,d(w_1, w_2) <
\varepsilon')))$, where $|w|_I$ denotes the integer part length of the
encoding $w$.

By hypothesis, there exists $x_1 \in S$ such that $\forall
\varepsilon_1 \exists x_2 \forall \varepsilon_2 \exists x_3 \forall
\varepsilon_3 \cdots$, $d(x_i, x_{i+1}) \leq \varepsilon_i$ for all $i
\geq 1$, $x_i \in S$ for all odd $i$, and $x_i \notin S$ for all even
$i$. We choose $S_1 = S$, and define $S_2$ as the subset of
$\overline{S}$ whose elements $x_2$ satisfy $\forall \varepsilon_2
\exists x_3 \forall \varepsilon_3 \exists x_4 \forall \varepsilon_4
\cdots$, $d(x_i, x_{i+1}) \leq \varepsilon_i$ for all $i \geq 2$, $x_i
\in S$ for all odd $i$, and $x_i \notin S$ for all even $i$. By the
previous property, there exists an encoding $w_1$ of $x_1$ such that
for arbitrarily small $\varepsilon' > 0$, there exists an element
$x_2$ of $S_2$ whose all encodings $w_2$ satisfy $d(w_1, w_2) <
\varepsilon'$, provided that they share the same integer-part length
as $w_1$. Moreover, there exists at least one such encoding $w_2$. By
applying a similar reasoning to $x_2, x_3, x_4, \ldots$, one obtains
$\exists w_1\forall \varepsilon'_1 \exists w_2 \forall \varepsilon'_2
\exists w_3 \forall \varepsilon'_3 \cdots$, $d(w_i, w_{i+1}) \leq
\varepsilon'_i$ for all $i \geq 1$, $w_i \in L_r(S)$ for all odd $i$,
and $w_i \notin L_r(S)$ for all even $i$. It follows that the language
$L_r(S)$ satisfies the dense oscillating sequence property.  \qed 

\subsubsection{Links between the topology of $\omega$-words and 
the topology of real numbers}

In this section, the notations $F_\sigma$, $G_\delta$ and 
$F_\sigma \cap G_\delta$ (resp. ${\sf F_\sigma}$,
${\sf G_\delta}$ and ${\sf F_\sigma \cap G_\delta}$) will be used when
dealing with the topology of $\omega$-words (resp. real numbers).

\begin{lem}
\label{lemma-f-sigma}
Let $r \in \nats_{>1}$ be a base, and let $L \subseteq (\Sigma_r \cup
\{\star\})^\omega$ be a language.  If $L$ belongs to $F_\sigma$, 
then the set of real numbers that have an encoding in $L$ belongs to 
${\sf F_\sigma}$.
\end{lem}

\proof\
Let $W_{j}$ be the
language $\{0,r-1\}\{0,\ldots,r-1\}^j\star(\Sigma_r \cup \{\star\})^
\omega$ with $j \in \nats$.  This language is open for all $j$.  
Since $L$ belongs to $F_\sigma$, it can be expressed as 
$L = \bigcup_{i \in \nats}F_i$, where each $F_i$ is closed.  The language
$\bigcup_{i \in \nats}\bigcup_{j \in \nats}(F_i \cap W_{j})$ 
is a sublanguage of $L$ such that the language of valid encodings it contains
is exactly the language of valid encodings that belong to $L$.
When $i$ and $j$ are fixed, the set $F_i \cap W_{j}$ is the
intersection of a closed and an open set; hence, it belongs to
$F_\sigma$ and is thus a countable union of closed sets: 
$F_i \cap W_{j} = \bigcup_{k \in \nats} L_{i,j,k}$.  

For each of these closed sets $L_{i,j,k}$, define $S_{i,j,k} \subseteq
\reals$ as the set of numbers that have at least one encoding in
$L_{i,j,k}$.  The set $S_{i,j,k}$ is closed. Indeed, suppose that
$S_{i,j,k}$ is not closed. Thus, there exists a converging sequence of
points of $S_{i,j,k}$ whose limit $x$ does not belong to $S_{i,j,k}$.
If this sequence contains infinitely many points greater than $x$, one
extracts its subsequence composed of those points.  Otherwise, one
extracts the subsequence composed of its points that are lower than $x$. Each
of the points of $S_{i,j,k}$ has at least one encoding in $F_i \cap
W_{j}$.  Since the valid encodings in $F_i \cap W_{j}$ have the same
integer part length, the converging subsequence of points of
$S_{i,j,k}$ is mapped to a converging sequence of words encoding those
points.  Since $L_{i,j,k}$ is closed, it contains the limit of its
converging sequences, hence the limit $x$ of the converging sequence
of points of $S_{i,j,k}$ has an encoding in $L_{i,j,k}$, which leads
to a contradiction since this limit would be in $S_{i,j,k}$.

It follows that the set of real numbers that have an encoding in $L$
is a countable union $\bigcup_{(i,j,k) \in \nats^3}S_{i,j,k}$ of
closed sets in $\reals$, and thus belongs to ${\sf F_\sigma}$. 
\qed

\begin{lem}
\label{lemma-f-sigma-g-delta}

Let $S \subseteq \reals$, and $r \in \nats_{>1}$ be a base.
The set $S$ belongs to ${\sf F_\sigma \cap G_\delta}$ 
iff the language $L_r(S)$ belongs to
$F_\sigma \cap G_\delta$.
\end{lem}

\proof\ 
It is known~\cite{BJW05} that if a set $S \subseteq \reals$ belongs to 
${\sf F_\sigma \cap G_\delta}$, 
then the language $L_r(S)$ belongs to $F_\sigma \cap G_\delta$.

If $L_r(S)$ belongs to $F_\sigma \cap G_\delta$, 
then it belongs in particular to $F_\sigma$.  
By Lemma~\ref{lemma-f-sigma}, $S$ then belongs to ${\sf F_\sigma}$.  
On the other hand, $L_r(S)$ belongs
to $G_\delta$.  It follows that the complement of $L_r(S)$ belongs to 
$F_\sigma$.  
By Lemma~\ref{lemma-f-sigma}, the set of real numbers that have
an encoding in this language belongs to ${\sf F_\sigma}$, 
which implies that $S$ belongs to ${\sf G_\delta}$.
\qed

In the sequel, we will need to apply
transformations to sets represented by RNA (or weak RNA), or to the 
chosen encoding base.

\begin{thm}
\label{theo-affine}
Let $S \subseteq \reals$, $r \in \nats_{>1}$, and $a, b\in \rationals$. 
If $S$ is (resp. weakly) $r$-recognizable
then the sets $aS + b$ and 
$S \cap [a, b]$ are (resp. weakly) $r$-recognizable as well.
\end{thm}

\proof\ If $S$ is $r$-recognizable, then it is definable in $\langle
\reals, \ints, +, <, X_r \rangle$ by Theorem~\ref{theo-brw98}, and so
are the sets $aS + b$ and $S \cap [a, b]$, that thus are both
$r$-recognizable.

If $S$ is weakly $r$-recognizable, 
then the language $L_r(S)$ belongs
to the class $F_\sigma \cap G_\delta$.
By Lemma~\ref{lemma-f-sigma-g-delta}, the set $S$ belongs to the class
${\sf F_\sigma \cap G_\delta}$, and so are the
sets $aS + b$ and $S \cap [a, b]$.
Since these sets are $r$-recognizable by the first part of the proof,
it follows from~\cite{MS97} that they are also weakly $r$-recognizable.
\qed

\begin{thm}
\label{theo-base-power}
Let $S \subseteq \reals$, $r \in \nats_{>1}$, and $l \in \nats_{>0}$. The
set $S$ is (resp. weakly) $r$-recognizable iff it is
(resp. weakly) $r^l$-recognizable.
\end{thm}

\proof\ If $S$ is $r$-recognizable, then the result is a consequence
of Theorem~\ref{theo-brw98}, since the predicate $X_r(x,u,k)$ can be
expressed in terms of $X_{r^l}(x,u,k)$, and reciprocally.  Indeed,
testing the value of the digit at a given position in an encoding in
base $r^l$ can be reduced to the test of $l$ digits in base $r$, and
conversely.

If $S$ is weakly $r$-recognizable, then $L_r(S)$ belongs to the class 
$F_\sigma \cap G_\delta$.  By
Lemma~\ref{lemma-f-sigma-g-delta}, $S$ belongs to the class
${\sf F_\sigma \cap G_\delta}$, and $L_{r^l}(S)$
belongs to the class $F_\sigma \cap G_\delta$.  Since
$S$ is $r^l$-recognizable, $S$ is weakly 
$r^l$-recognizable.  The case of a $r^l$-recognizable set $S$ is
handled in the same way.
\qed

\section{Prior results and objectives}

This article is aimed at characterizing precisely the conditions under
which a set of real numbers is recognizable, or weakly recognizable,
in multiple bases. We start by summarizing some known results.

First, the case of sets of integer numbers is handled by the following
result, which is a direct corollary of the well-known {\em Cobham's
  theorem\/}. Note that for sets of integer
numbers, the notions of $r$-recognizability and weak
$r$-recognizability coincide, and correspond to the existence of
a finite-word automaton accepting only the integer part of encodings.

\begin{thm}[\cite{Cob69,BHMV94}]
\label{theo-cobham}
Let $r, s \in \nats_{>1}$ be bases that are {\em multiplicatively
  independent}, i.e., such that $r^p \neq s^q$ for all $p,q \in
\nats_{>0}$. A set $S \subseteq \ints$ is both $r$- and
$s$-recognizable iff it is definable in the first-order theory
$\langle \ints, +, < \rangle$.
\end{thm}

If $r, s \in \nats_{>1}$ are multiplicatively dependent, then a set $S
\subseteq \ints$ is $r$-recognizable iff it is $s$-recognizable, as a
consequence of Theorem~\ref{theo-base-power}. It follows that
Theorem~\ref{theo-cobham} fully characterizes recognizability in
multiple bases for sets of integer numbers.

Next, for sets of real numbers recognized by weak automata, we have
the following result.

\begin{thm}[\cite{BB09}]
\label{theo-bb07}
Let $r, s \in \nats_{>1}$ be bases that do not share the same set
of prime factors. A set $S \subseteq \reals$ is both weakly $r$- and
weakly $s$-recognizable iff it is definable in the first-order theory
$\langle \reals, \ints, +, < \rangle$.
\end{thm}

In this paper, we extend Theorem~\ref{theo-bb07} in two ways.  First,
we will show in Section~\ref{sec-mult-indep-bases} that this result
also holds for multiplicatively independent bases, which weakens the
hypotheses of the theorem and brings its statement closer to
Theorem~\ref{theo-cobham}. Formally, we will prove the following
theorem.

\begin{thm}
\label{theo-cobham-reals} 
Let $r, s \in \nats_{>1}$ be two multiplicatively independent bases.  A set $S
\subseteq \reals$ is both weakly $r$- and weakly $s$-recognizable iff
it is definable in the first-order theory $\langle \reals, \ints, +,
<\rangle$.
\end{thm}

Second, we will establish in
Section~\ref{sec-diff-prime-sets} that a similar result holds for
recognizable (as opposed to weakly recognizable) sets of real
numbers. Formally, we will prove the following theorem.

\begin{thm}
\label{theo-cobham-reals-general}
Let $r, s \in \nats_{>1}$ be two bases that do not share the same set
of prime factors.  A set $S \subseteq \reals$ is both r- and
s-recognizable iff it is definable in the first-order theory $\langle
\reals, \ints, +, <\rangle$.
\end{thm}

In this particular case, we will also show that considering bases with
different sets of prime factors is essential, and that
multiplicatively independent bases do not lead to a similar property.

Before proving Theorems~\ref{theo-cobham-reals}
and~\ref{theo-cobham-reals-general}, we show in the next section that
these problems can be reduced to simpler ones.

\section{Problem reductions}

In the next sections, we will consider sets $S \subseteq \reals$ that
are simultaneously recognizable, either by RNA or by weak RNA, in two
bases $r$ and $s$ that either are multiplicatively independent, or
have different sets of prime factors.  We will then tackle the problem
of proving that such sets are definable in $\langle \reals, \ints, +,
<\rangle$. In this section, we reduce this problem, by restricting the
domain to the interval $[0, 1]$, and introducing the notion of
boundary point.

\subsection{Reduction to $[0,1]$}
\label{sec-reduction-interval}

This section is adapted from~\cite{BB09}.
Let $S \subseteq \reals$ be a set of real numbers.  The set $S$ can be
decomposed into a countable union 
\[\bigcup_{i \in \ints} (\{i\} + S_i^F),\]
where for all $i$, $S_i^F \subseteq [0,1]$ is the set of fractional parts
that can be added to the integer $i$ to obtain an element $x \in S$.

If we decompose the set $\ints$ into equivalence classes $S_1^I,
S_2^I, S_3^I, \ldots$ such that
two integers $i$ and $j$ are in the same equivalence class iff 
the sets $S_i^F$ and $S_j^F$ are identical, then this union becomes a
(finite or infinite) union
\[\bigcup_{i}(S_i^I + S_i^F).\]

Assume now that $S$ is recognizable by a (resp. weak) RNA ${\cal A}$
in some base $r \in \nats_{>1}$. Recall that ${\cal A}$ has a
deterministic transition relation.  For each encoding of each possible
value $x_I \in \ints$ , the path in ${\cal A}$ that reads this
encoding followed by the separator $\star$ leads to a state $q$
accepting a language $L_q$.  The language $0^+ \star L_q$ encodes the
set of all fractional parts $x_F$ that can be associated to $x_I$,
i.e., the set $S_q = \{x_F \in [0,1] \mid x_I + x_F \in S\}$.  Note
that the dual encodings of $0$ and $1$ may be missing, but this is not
problematic.

Such states $q$ are in a finite number $n$, and can w.l.o.g. be
supposed to accept languages that are pairwise different (otherwise,
it suffices to modify the destinations of the transitions labeled by
$\star$ that lead to redundant states).  Assuming w.l.o.g. that the
languages accepted from every state are not empty, it follows that the
languages $L_q$ are in the same finite number $n$, and so are the sets
$S_q$.  The sets $S_q$ correspond exactly to those of the sets $S_i^F$
that are not empty.  Hence, the number $n$ of sets $S_q$
is independent from the representation base.

The set $S$ can thus be decomposed into a finite union
\[\bigcup_{i = 1}^n (S_i^I + S_i^F),\]
where the sets $S_i^I\subseteq \ints$ are non-empty and pairwise
distinct, and the sets $S_i^F \subseteq [0,1]$ are non-empty and
pairwise different.  Furthermore, each set $S_i^I$ is recognizable by
a finite-word automaton in every base in which $S$ is recognizable,
and each set $S_i^F$ is (resp. weakly) recognizable in every base in
which $S$ is (resp. weakly) recognizable\footnote{Indeed, in any
  (resp. weak) RNA recognizing the set $S$, there exists a state $q$
  accepting a language $L_q$ such that the language encoding $S_i^F$
  is $0^+\star L_q$.  In order for such a language to contain all
  encodings of the numbers it encodes, it should also contain the
  words $(r-1)^+ \star (r-1)^\omega$ if $0^\omega \in L_q$, and
  $0^+1\star 0^\omega$ if $(r-1)^\omega \in L_q$.}.

Assume now that $S \subseteq \reals$ is simultaneously (resp. weakly)
$r$- and $s$- recognizable, with respect to bases $r$ and $s$ that
are multiplicatively independent.  By Theorem~\ref{theo-cobham},
each set $S_i^I$ is thus definable in $\langle \ints, +, < \rangle$.
This reduces the problem of establishing that $S$ is definable in
$\langle \reals, \ints, +, < \rangle$ to the same problem for each
set $S_i^F$.  Since we have $S_i^F \subseteq [0,1]$ for all $i$,
the problem has thus been reduced from the domain $\reals$ to
the interval $[0,1]$.

\subsection{Boundary points}
\label{sec-reduction-boundaries}

A point $x \in \reals$ is a {\em
  boundary point\/} of a set $S \subseteq \reals$ iff all its
neighborhoods contain at least one point from $S$ as well as one from its
complement $\overline{S} = \reals \setminus S$.

\begin{lem} \label{lemma-set-boundary-points}
Let $r \in \nats_{>1}$ be a base.
If a set $S \subseteq \reals$ is $r$-recognizable, then
the set $B_S$ of boundary points of $S$ is $r$-recognizable.
\end{lem}

\proof\ Since $S$ is $r$-recognizable, it is definable in $\langle
\reals, \ints, +, <, X_r \rangle$ by Theorem~\ref{theo-brw98}. It is
sufficient to show that $B_S$ is definable in $\langle \reals, \ints,
+, <, X_r \rangle$.  A formula defining $B_S$ in this theory is
\[
\{x \in \reals \mid (\forall \varepsilon \in \reals_{>0})
(\exists y,z \in \reals) (y \in S \wedge z \notin S \wedge |x - y | <
\varepsilon \wedge |x-z| < \varepsilon\}.
\]
\qed

\begin{lem}
\label{lemma-finite-boundary}
Let $r \in \nats_{>1}$ be a base.
If a $r$-recognizable set $S \subseteq \reals$ has only finitely many
boundary points, then it is definable in the first-order theory
$\langle \reals, \ints, +, <\rangle$.
\end{lem}

\proof\ If $S \subseteq \reals$ has only finitely many boundary
points, then it can be decomposed into a finite union of intervals such
that the extremities of these intervals are the boundary points of $S$.

In order to prove that $S$ is definable in $\langle \reals, \ints, +,
<\rangle$, it is sufficient to show that the boundary points of $S$
are rational numbers. Since $S$ is $r$-recognizable,
the finite set $B_S$ of its boundary points is $r$-recognizable by Lemma
\ref{lemma-set-boundary-points}. It follows that its elements are encoded
by words accepted by a finite automaton, and that share a finite number
of fractional parts. These are necessarily ultimately periodic, 
from which the elements of $B_S$ are rational.
\qed

\section{Multiplicatively independent bases}
\label{sec-mult-indep-bases}

Let $r, s \in \nats_{>1}$ be two multiplicatively independent
bases. The first aim of this section is to prove
Theorem~\ref{theo-cobham-reals}, i.e., to establish that the subsets of
$\reals$ that are both weakly $r$- and weakly $s$-recognizable are
exactly those that are definable in $\langle \reals, \ints, +, <
\rangle$. Then, a second goal will be to show that the subsets of
$\reals$ that are both $r$- and $s$-recognizable do not enjoy the same
property.

Thanks to the reduction discussed in
Section~\ref{sec-reduction-interval}, it is sufficient to prove these
results for sets restricted to the interval $[0, 1]$. Besides,
Lemma~\ref{lemma-finite-boundary} implies that, in order to show that a
recognizable set is definable in $\langle \reals, \ints, +, <\rangle$,
it suffices to prove that it admits only a finite number of boundary
points.

We thus proceed as follows. We consider a set $S \subseteq [0, 1]$
that is both (resp. weakly) $r$- and $s$-recognizable, and assume by
contradiction that $S$ has infinitely many boundary points.  In
Section~\ref{sec-prod-stab}, we derive some useful properties under
this assumption. In Section~\ref{sec-recog-weak}, we then show that
our assumption leads to a contradiction in the case of weak
recognizability, proving that the sets that are both weakly $r$- and
weakly $s$-recognizable are necessarily definable in $\langle \reals,
\ints, +, < \rangle$, hence Theorem~\ref{theo-cobham-reals}. Finally,
in Section~\ref{sec-recog-rna}, we show with the help of a
counterexample that this result does not generalize to sets that are
both $r$- and $s$-recognizable.

\subsection{Product stability}
\label{sec-prod-stab}

By hypothesis, the set $S \subseteq [0, 1]$ is (resp. weakly)
$r$-recognizable.  Let ${\cal A}_r$ be a (resp. weak) RNA recognizing
$S$ in base $r$. We assume w.l.o.g. that the transition relation of
${\cal A}_r$ is complete.

Since $S$ is $r$-recognizable, the set $B_S$ of boundary
points of $S$ is $r$-recognizable by Lemma~\ref{lemma-set-boundary-points}. 
Let ${\cal A}_r^B$ be a RNA recognizing $B_S$.

By assumption, $S$ has infinitely many boundary points, hence
there exist infinitely many distinct paths of ${\cal A}_r^B$ that end
up cycling in the same set of accepting states.
One can thus extract
from ${\cal A}_r^B$ an infinite language $L = 0 \star u v^* t w^{\omega}$, 
where $t, u, v, w \in \Sigma_r^*$, $|v| > 0$, $|w| > 0$, 
and $L$ encodes an infinite subset of the boundary
points of $S$. We then define $y = [0 \star u v^{\omega}]_r$ and, for
each $k \in \nats_{>0}$, $y_k = [0 \star u v^k t w^{\omega}]_r$. The
sequence $y_1, y_2, y_3, \ldots \in \rationals^{\omega}$ forms an
infinite sequence of distinct boundary points of $S$, converging
to $y \in \rationals$. If we have $y_k > y$ for infinitely many
$k$, then we define $S^{1} = (S - y) \cap [0, 1]$. Otherwise, we
define $S^{1} = (-S + y) \cap [0, 1]$. From Theorem~\ref{theo-affine},
the set $S^{1}$ is both (resp. weakly) $r$- and $s$-recognizable. 
Moreover, this set
admits an infinite sequence of distinct boundary points that converges
to $0$.

Let ${\cal A}^1_r$ and ${\cal A}^1_s$ be (resp. weak) RNA
recognizing $S^{1}$ in the respective bases $r$ and~$s$. The path
$\pi_0$ of ${\cal A}^1_r$ that reads $0 \star 0^{\omega}$ is composed of 
a prefix labeled by $0 \star$, followed by an acyclic path of
length $p \geq 0$, and finally by a cycle of length $q > 0$. It
follows that a word of the form $0 \star 0^p t$, with $t \in
\Sigma_r^{\omega}$, is accepted by ${\cal A}^1_r$ iff the word $0 \star
0^{p+q} t$ is accepted as well. Remark that the set $S^{1}$ admits
infinitely many boundary points with a base-$r$ encoding beginning
with $0 \star 0^p$.  Similar properties hold for ${\cal A}^1_s$. In this
automaton, the path $\pi'_0$ recognizing $0 \star 0^{\omega}$ reads the
symbols $0$ and $\star$, and then follows an acyclic sequence of length $p'$
before reaching a cycle of length $q'$.

We now define $S^{2} = r^p S^{1}\cap [0, 1]$. Like $S^{1}$, the set
$S^{2}$ admits an infinite sequence of boundary points that converges
to $0$. Moreover, by Theorem~\ref{theo-affine}, $S^{2}$ is both (resp. weakly) 
$r$- and $s$-recognizable.
Let ${\cal A}^2_r$ be a (resp. weak) RNA recognizing $S^{2}$
in base $r$. For every $t \in \Sigma_r^{\omega}$, the word $0\star t$
is accepted by ${\cal A}^2_r$ iff the word $0\star 0^q t$ is accepted
as well. In other words, the fact that a number $x \in [0, 1]$ belongs
or not to $S^{2}$ is not influenced by the insertion of $q$ zero
digits in its encodings, immediately after the symbol $\star$. This
amounts to dividing the value of $x$ by $r^q$, which leads to the
following definition.

\begin{defi}
Let $D \subseteq \reals$ be a domain, and let $f \in \reals_{>0}$. A
set $S \subseteq D$ is\/ {\em $f$-product-stable\/} in the domain $D$
iff for all $x \in D$ such that $fx \in D$, we have $x \in S
\Leftrightarrow fx \in S$.
\end{defi}

{}From the previous discussion, we have that $S^{2}$ is
$r^q$-product-stable in $[0, 1]$. We then define $S^{3} = s^{p'}S^{2}
\cap [0, 1]$. The set $S^{3}$ is $r^q$-product-stable in $[0, 1]$ as
well. By Theorem~\ref{theo-affine}, $S^{3}$ is also both (resp. weakly)
$r$- and $s$-recognizable.
Besides, since $S^{3} = r^p s^{p'} S^{1} \cap [0,
  1]$, the set $S^{3}$ can alternatively be obtained by first defining
$S^{4} = s^{p'} S^{1}\cap [0, 1]$, which is both (resp. weakly) $r$- and
$s$-recognizable by Theorem~\ref{theo-affine}. Then, one has $S^{3} =
r^{p}S^{4} \cap [0, 1]$.  By a similar reasoning in base $s$, we get
that $S^{3}$ is $s^{q'}$-product-stable in $[0, 1]$. Like $S^{2}$, the
set $S^{3}$ admits an infinite sequence of distinct boundary points
that converges to $0$.

Finally, we replace the bases $r$ and $s$ by $r' = r^q$ and $s' =
s^{q'}$, thanks to Theorem~\ref{theo-base-power}. The results of this
section are then summarized by the following lemma.

\begin{lem}
\label{lemma-stability}
Let $r, s \in \nats_{>1}$ be two multiplicatively independent bases, and let $S
\subseteq [0, 1]$ be a set that is both (resp. weakly) $r$- and 
$s$-recognizable, and that admits infinitely many boundary points. 
There exist powers $r' = r^i$ and $s' = s^j$ of $r$ and $s$, with 
$i, j \in \nats_{>0}$, and a set $S' \subseteq [0, 1]$ that is both 
(resp. weakly) $r'$- and $s'$-recognizable, 
both $r'$- and $s'$-product-stable 
in $[0, 1]$, and that admits infinitely many boundary points.
\end{lem}

\subsection{Recognizability by weak RNA}
\label{sec-recog-weak}

We are now ready to prove that our initial assumption that the set $S
\subseteq [0, 1]$
has infinitely many boundary points leads to a contradiction, under
the hypothesis that $S$ is both weakly $r$- and weakly
$s$-recognizable.

By Lemma~\ref{lemma-stability}, we can assume w.l.o.g.  that $S$ is
$r$- and $s$-product-stable in $[0, 1]$.  Hence, there exist $\alpha,
\beta \in (0, 1]$ such that $\alpha \in S$ and $\beta \not\in S$. For
  every $i, j \in \ints$ such that $r^i s^j \alpha \in (0, 1]$, we
    thus have $r^i s^j \alpha \in S$. Similarly, for every $i, j \in
    \ints$ such that $r^i s^j \beta \in (0, 1]$, we have $r^i s^j
      \beta \not\in S$.

Let $\gamma$ be an arbitrary point in the open interval $(0, 1)$. 
Since $r$ and $s$ are multiplicatively independent, it follows from
Kronecker's approximation theorem~\cite{HW85} that
any open interval of $\reals_{>0}$ contains some number of the form
$r^i/s^j$ with $i,j \in \nats_{> 0}$~\cite{Per90}. Hence, for every 
sufficiently small $\varepsilon > 0$ and $\delta \in \{ \alpha, \beta\}$, 
there exist $i, j \in \nats_{> 0}$ such that 
\[ 
0 < \gamma - \varepsilon < (r^i/ s^j) \delta < \gamma + \varepsilon < 1
\]
showing that every neighborhood $N_\varepsilon(\gamma)$ 
of $\gamma$ contains one point from $S$ as 
well as one from $\overline{S}$.
The latter property leads to a contradiction, since it
implies that $S$ satisfies the dense oscillating sequence property, and
therefore, by Theorem~\ref{thm-dense-oscillating}, cannot be recognized by a 
weak RNA.

Taking into account the problem reductions introduced in
Sections~\ref{sec-reduction-interval}
and~\ref{sec-reduction-boundaries}, we thus have proven
Theorem~\ref{theo-cobham-reals}.

Thanks to the above mentioned reductions,
Theorem~\ref{theo-cobham-reals} has the following corollary. A set $S
\subseteq \reals$ is weakly $r$- and weakly $s$-recognizable in two
multiplicatively independent bases iff it can be expressed as a finite
union $\bigcup_i (S_i^I + S_i^F)$, where each $S_i^I \subseteq \ints$
is of the form $S^I_i = \{ a_i + kb_i \mid k \in \nats \}$ with $a_i,
b_i \in \ints$, and each $S_i^F \subseteq [0, 1]$ is a finite union of
intervals with rational extremities.  It has already been observed
in~\cite{Wei99} that such a structural description of subsets of
$\reals$ is equivalent to definability in $\langle \reals, \ints, +,
<\rangle$.

\subsection{Recognizability by RNA}
\label{sec-recog-rna}

We now show that Theorem~\ref{theo-cobham-reals} does not directly
generalize to non-weak recognizability.  Indeed, a set can then be
recognizable in two multiplicatively independent bases without being
definable in $\langle \reals, \ints, +, < \rangle$. This property is
established by the following theorem.

\begin{thm}
For every pair of bases $r, s \in \nats_{>1}$ that share the same set
of prime factors, there exists a set $S$ that is both $r$- and
$s$-recognizable, and that is not definable in the first-order theory
$\langle \reals, \ints, +, < \rangle$.
\end{thm}

\proof\ A counterexample is provided by the set \[ S = \left\{ 
\frac{n}{f_1^{i_1}f_2^{i_2} \cdots f_k^{i_k}} 
\mid n \in \ints,\, i_1, i_2, \ldots, i_k \in \nats\right\}, \] 
where $f_1, f_2, \ldots f_k$ are the prime factors of
$r$ and $s$. 

In either base $t \in \{ r, s \}$, this set is
encoded by the language $L_t = \{0,t-1\} \Sigma_t^*\star \Sigma_t^*
(0^{\omega} \cup (t-1)^{\omega})$, i.e., the set $S$ contains the numbers
that admit dual encodings.  
Indeed, each word of $L_t$ represents a number $x = n/t^k$ 
($n \in \ints$, $k \in \nats$) that belongs to $S$.  Reciprocally, let
$x$ be an element of $S$.  One can assume w.l.o.g. that the denominator
of $x$ is a power of $t$.  Hence, $x$ admits an encoding that ends with
$0^\omega$.

The language $L_t$ is clearly $\omega$-regular, hence $S$ is both $r$-
and $s$-recognizable. Suppose that $S$ is definable in $\langle
\reals, \ints, +, < \rangle$.  Then, it is weakly $t$-recognizable in
any base $t$ thanks to Theorem~\ref{theo-bjw05}.  By
Theorem~\ref{thm-dense-oscillating} and since $S$ satisfies the dense
oscillating sequence property, this leads to a contradiction.  \qed

Note that the set $S$ (resp. $\reals \backslash S$) defined in the
previous proof is recognizable by deterministic co-B\"uchi automata
(resp. deterministic B\"uchi automata) in both bases $r$ and $s$.  It
follows that Theorem~\ref{theo-cobham-reals} does not generalize to
sets recognizable by those classes of automata either.

The case of bases that do not share the same set of prime factors is
investigated in the next section.

\section{Bases with different sets of prime factors}
\label{sec-diff-prime-sets}

We now consider two bases $r, s \in \nats_{>1}$ that do not share
the same set of prime factors. Since this property implies that $r$
and $s$ are multiplicatively independent, we know by 
Theorem~\ref{theo-cobham-reals} that any subset of $\reals$ that is
simultaneously weakly $r$- and weakly $s$-recognizable must be
definable in $\langle \reals, \ints, +, <\rangle$.

The goal of this section is now to prove
Theorem~\ref{theo-cobham-reals-general}, i.e., that a subset of
$\reals$ that is both $r$- and $s$-recognizable is necessarily
definable in $\langle \reals, \ints, +, <\rangle$. Recall that, as
shown in Section~\ref{sec-recog-rna}, this result does not extend to
pairs of bases that are multiplicatively independent but share the
same prime factors.

We proceed like in Section~\ref{sec-mult-indep-bases} and start from
an arbitrary set $S$ that is both $r$- and $s$-recognizable. Thanks to
the reduction discussed in Section~\ref{sec-reduction-interval}, it
suffices to consider $S \subseteq [0, 1]$. Moreover, according to
Lemma~\ref{lemma-finite-boundary}, one can prove that $S$ is definable
in $\langle \reals, \ints, +, <\rangle$ by showing that it admits only
finitely many boundary points. We thus assume that $S$ admits
infinitely many boundary points. From this assumption, we will derive
in Sections~\ref{sec-sum-stab} and~\ref{sec-expl-sum-stab} additional
properties that will eventually lead to a contradiction.

It is possible to reuse part of the reasoning made in
Section~\ref{sec-mult-indep-bases}.  By Lemma~\ref{lemma-stability},
there exist bases $r'$ and $s'$ with different sets of prime factors,
and a set $S' \subseteq [0, 1]$ that is both $r'$- and
$s'$-recognizable, both $r'$- and $s'$-product-stable in $[0, 1]$, and
that has infinitely many boundary points. Replacing the set $S$ by
$S'$, and the bases $r, s$ by $r', s'$, we can thus assume
w.l.o.g. that the set $S$ that we consider is both $r$- and
$s$-product-stable in $[0, 1]$. Finally, we also impose w.l.o.g. that
there exists a prime factor of $s$ that does not divide $r$.

\subsection{Sum stability}
\label{sec-sum-stab}
 
Our first strategy consists in exploiting Cobham's theorem so as to
derive additional properties of $S$. The initial step is to build from
$S$ a set $S' \subseteq \reals_{\geq 0}$ that coincides with $S$ over
$[0, 1]$, shares the same recognizability and product-stability
properties, and contains numbers with non-trivial integer parts.

\begin{lem}
\label{lemma-wroclaw-constr}
Let $r,s \in \nats_{>1}$ be two bases with different sets of prime
factors, and let $S \subseteq [0, 1]$ be a set that is 
$r$- and $s$-recognizable,
$r$- and $s$-product-stable in $[0, 1]$, and that has infinitely many
boundary points.  There exists a set $S' \subseteq \reals_{\geq 0}$
that is $r$- and $s$-recognizable, $r$- and
$s$-product-stable in $\reals_{\geq 0}$, and that has infinitely many
boundary points.
\end{lem}
\proof\ Let $S' = \{ r^k x \mid x \in S \wedge k \in \nats \}$. This
set is clearly $r$-product-stable in $\reals_{\geq 0}$.  Since $S$ is
$r$-product-stable in $[0, 1]$, we have $S' \cap [0, 1] = S$ showing 
that $S'$ has infinitely many boundary points. 
A RNA ${\cal A}'_r$ recognizing $S'$ in base $r$ is built from
an automaton ${\cal A}_r$ recognizing $S$ by delaying 
arbitrarily the reading of the symbol $\star$.  In other words, 
a word $u v \star w$ is accepted by ${\cal A}'_r$, with
$u \in \{0,r-1\}\Sigma_r^*$, $v \in \Sigma_r^*$, 
and $w \in \Sigma_r^\omega$, whenever the word $u \star v w$ is accepted by 
${\cal A}_r$.

In order to prove that $S'$ is $s$-recognizable, notice that, since
$S$ is both $r$- and $s$-product-stable in $[0, 1]$, we have $S' = \{
r^i s^j x \mid x \in S \wedge i, j \in \ints \}$. The set $S'$ can
therefore be expressed as $S' = \{ s^k x \mid x \in S \wedge k \in
\nats \}$.  By the same reasoning as in base $r$, this set is
$s$-recognizable, as well as $s$-product-stable in $\reals_{\geq 0}$. \qed

Consider now a set $S'$ obtained from $S$ by
Lemma~\ref{lemma-wroclaw-constr}.  As discussed in
Section~\ref{sec-reduction-interval}, this set can be expressed as a
finite union $S' = \bigcup_i (S_i^I + S_i^F)$, where for each $i$, we
have $S_i^I \subseteq \nats$ and $S_i^F \subseteq [0, 1]$. Moreover,
for each $i$, the set $S_i^I$ is both $r$- and $s$-recognizable, and
it follows from Theorem~\ref{theo-cobham} that this set is definable
in $\langle \nats, +, < \rangle$.  Since such a set is ultimately
periodic~\cite{Cob69,BHMV94}, there exists $n_i \in \nats_{>0}$ for
which $\forall x \in \nats,\, x \geq n_i :\, x \in S_i^I
\Leftrightarrow x + n_i \in S_i^I$.  By defining $n = \lcm_i(n_i)$, we
obtain $\forall x \in \reals_{\geq 0},\, x \geq n :\, x \in S'
\Leftrightarrow x + n \in S'$. This prompts the following definition.

\begin{defi}
Let $D \subseteq \reals$ be a domain, and let
$t \in \reals$. A set $S \subseteq D$ is\/ {\em $t$-sum-stable\/}
in $D$ iff for all $x \in D$ such that $x+t \in D$,
we have $x \in S \Leftrightarrow x + t \in S$.
\end{defi}

Let us show that the set $S'' = (1/n)S' \backslash \{0\}$ is $1$-sum-stable in
$\reals_{> 0}$. For every $x \geq 1$, we have $x \in S''
\Leftrightarrow x + 1 \in S''$. For $x < 1$, we choose $k \in \nats$
such that $r^k x \geq 1$. Exploiting the properties of $S'$
(transposed to $S''$), we get $x \in S'' \Leftrightarrow r^k x \in S''
\Leftrightarrow r^k x + r^k \in S'' \Leftrightarrow x + 1 \in
S''$. 
Lemma~\ref{lemma-wroclaw-constr} can thus be refined as follows.

\begin{lem}
\label{lemma-main}
Let $r,s \in \nats_{>1}$ be two bases with different sets of prime
factors, and let $S \subseteq [0,1]$ be a set that is $r$- and 
$s$-recognizable, $r$- and $s$-product-stable in $[0,1]$, 
and that has infinitely many boundary points.  There exists a set $S'
\subseteq \reals_{>0}$ that is $r$- and $s$-recognizable, has
infinitely many boundary points, and is $r$-product-, $s$-product- 
and $1$-sum-stable in $\reals_{>0}$.
\end{lem}

Note that Lemmas \ref{lemma-wroclaw-constr} and \ref{lemma-main} 
still hold if the bases $r$ and $s$ are multiplicatively independent.

\subsection{Exploiting sum-stability properties}
\label{sec-expl-sum-stab}

Consider a set $S' \subseteq \reals_{> 0}$ that satisfies the
properties expressed by Lemma~\ref{lemma-main}. It remains to show
that these properties lead to a contradiction.  
The hypothesis on the prime factors of $r$ and $s$ is explicitly
used in this section.

We proceed by characterizing the numbers $t \in \reals$ for which $S'$
is $t$-sum-stable in $\reals_{> 0}$. These form the set $T_{S'} = \{ t
\in \reals \mid \forall x \in \reals_{> 0} :\, x + t \in \reals_{>0}
\Rightarrow (x \in S' \Leftrightarrow x + t \in S') \}$. Since $S'$ is
$r$-recognizable, it is definable in $\langle \reals, \ints, +, <, X_r
\rangle$ by Theorem~\ref{theo-brw98}, and so is $T_{S'}$, that is
therefore $r$-recognizable as well.

The set $T_{S'}$ enjoys interesting closure properties: 

\begin{property}
\label{prop-closure}
For every $t,
u \in T_{S'}$ and $a, b \in \ints$, we have $at + bu \in T_{S'}$.
\end{property}

The set $T_{S'}$ is also $r$- and $s$-product stable in $\reals$. Since
$1 \in T_{S'}$, this yields the following property.

\begin{property}
\label{prop-digit-change}
For every $k \in \ints$, we have $r^k \in T_{S'}$ and  $s^k \in T_{S'}$.
\end{property}

Intuitively, being able to add or subtract $r^k$ from a number, for
any $k$, makes it possible to change in an arbitrary way finitely many
digits in its base-$r$ encodings, without influencing the fact that
this number belongs or not to $S'$. Our next step will be to show that
this property can be extended to all digits of base-$r$ encodings,
implying either $S' = \emptyset$ or $S' = \reals_{>0}$. This would then
contradict our assumption that $S'$ has infinitely many boundary
points.

\begin{lem} \label{lemma-lengths-periods}
Let $r,s \in \nats_{>1}$ be two bases such that $s$
has a prime factor that does not divide $r$.
The lengths of the smallest periods of the base-$r$ encodings
of $1/s^k$ are unbounded w.r.t. $k$.
\end{lem}

\proof\ 
The base-$r$ encodings of $1/s^k$ are of the form $0^+ \star v_k u_k^\omega$,
with $v_k \in \Sigma_r^*$ and $u_k \in \Sigma_r^+$.  We have
\[ \frac{r^{|v_k|}}{s^k} = [0v_k\star u_k^\omega]_r, \]
\[ \frac{r^{|v_k|+|u_k|}}{s^k} = [0v_ku_k \star u_k^\omega]_r.\]
Hence, 
\[ \frac{1}{s^k} = \frac{a_k}{r^{|v_k|}(r^{|u_k|}-1)},\]
with $a_k = [v_ku_k]_r - [v_k]_r \in \nats_{>0}$.

It follows that the lengths $|u_k|$ and $|v_k|$ are the smallest naturals
such that $s^k$ divides $r^{|v_k|}(r^{|u_k|}-1)$.  By hypothesis, there exists 
a prime factor $f$ of $s$ that does not divide $r$.  This implies that
the lengths of the periods $u_k$ must be unbounded w.r.t. $k$.
\qed

\begin{property}
\label{prop-period-change}
There exist $l, m\in \nats_{>0}$ such that, for every $k
\in \nats_{>0}$, we have \[ \frac{m}{r^{lk} - 1} \in T_{S'}. \]
\end{property}
\proof\ By Property~\ref{prop-digit-change}, we have $1/s^k \in T_{S'}$
for all $k \in \nats$. From Lemma~\ref{lemma-lengths-periods}, 
the lengths of the smallest  periods $u_k$ of the base-$r$ encodings 
of $1/s^k$ are unbounded w.r.k. $k$.

Consider a RNA ${\cal A}^T_r$ recognizing $T_{S'}$ in base $r$. We
study the rational numbers accepted by ${\cal A}^T_r$, which have
base-$r$ encodings of the form $v\star w u^{\omega}$.  We assume
w.l.o.g. that the considered periods $u$ are the shortest possible
ones.  It follows from the unboundedness of $u_k$ that $T_{S'}$
contains rational numbers with infinitely many distinct periods. 
RNA are deterministic Muller automata; hence, their accepting conditions
are finite unions of subsets of their set of states.  An infinite number
of encodings of rationals with distinct periods thus end in
exactly the same subset of accepting states.
In particular, there exist $u, u', v, v', w, w'$ such that $u^{\omega}$ is
not a suffix of $(u')^{\omega}$, the words $v\star wu^{\omega}$ and $v'\star
w'(u')^{\omega}$ are both accepted by ${\cal A}^T_r$, and the paths
$\pi$ and $\pi'$ of ${\cal A}^T_r$ reading them end up cycling in
exactly the same subset of accepting states.

Let $q$ be one of these states, and $u_1, u_2 \in \Sigma_r^+$ be
periods of the (respective) words read by $\pi$ and $\pi'$ after
reaching $q$ in their final cycle. These periods can be repeated
arbitrarily, hence we can assume w.l.o.g. that $|u_1| = |u_2|$. Moreover 
we can assume w.l.o.g.  that
$[u_2]_r > [u_1]_r$, otherwise $u^{\omega}$ would be
a suffix of $(u')^{\omega}$. Besides, there exist $v, w \in
\Sigma_r^*$ such that $v \star w$ reaches $q$. From the structure of
${\cal A}^T_r$, it follows that for every $k \geq 0$, the word $v
\star w (u_1^k u_2)^{\omega}$ is accepted by ${\cal A}^T_r$.

For each $k \geq 0$, we thus have $[v \star w (u_1^k
  u_2)^{\omega}]_r \in T_{S'}$. Developing, we get 
\[\frac{d_k + [vw\star 0^{\omega}]_r}{r^{|w|}}\in T_{S'},\] 
with $d_k = [\star (u_1^k u_2)^{\omega}]_r$. Thanks to
Properties~\ref{prop-closure} and \ref{prop-digit-change}, and the
$r$-product-stability property of $T_{S'}$, this implies $d_k \in
T_{S'}$. We now express $d_k$ in terms of $[u_1]_r$, $[u_2]_r$, and
$k$:
\[
d_k  =  \frac{[u_1^k u_2]_r}{r^{l(k + 1)} - 1}
 =  \frac{[u_2]_r - [u_1]_r}{r^{l(k + 1)} - 1} +
\frac{[u_1]_r}{r^l - 1}, \mbox{~where $l = |u_1| = |u_2|$.}
\]

The next step will consist in getting rid of the second term of this
expression. By Properties~\ref{prop-closure} and \ref{prop-digit-change}, 
we have for all $k \in \nats$,
\[
(r^l - 1)d_k - [u_1]_r =
\frac{m}{r^{l(k + 1)} - 1} \in T_{S'},
\]
where $m = (r^l - 1)([u_2]_r - [u_1]_r)$ is such that $m \in
\nats_{>0}$. For all $k > 0$, we thus have 
\[\frac{m}{r^{lk} - 1} \in T_{S'}.\eqno{\qEd}\]

We are now ready to conclude. Given $l$ and $m$ by
Property~\ref{prop-period-change}, we define $S'' = (1/m)S'$. Like $S'$,
this set has infinitely many boundary points.  The set $T_{S''}$ of the
values $t$ for which $S''$ is $t$-sum-stable in $\reals_{>0}$ is given
by $T_{S''} = (1/m) T_{S'}$. This set is thus $r$-recognizable.  From
Properties~\ref{prop-closure} and \ref{prop-digit-change}, 
we have for every $k \in \nats$, $1/r^k \in T_{S''}$.
Finally,
from Property~\ref{prop-period-change}, we have for every $k > 0$,
\[\frac{1}{r^{lk}- 1} \in T_{S''}.\]

\begin{property}
The set $T_{S''}$ is equal to $\reals$.
\end{property}
\proof\ Since $T_{S''}$ and $\reals$ are both $r$-recognizable, and two
$\omega$-regular languages are equal iff they share the same subset of
ultimately periodic words~\cite{PP04}, it is actually sufficient to
show that $T_{S''} \cap \rationals = \rationals$. Every rational $t$
admits a base-$r$ encoding of the form $v \star wu^{\omega}$, where
$|u| = lk$ for some $k \in \nats_{>0}$. We have 
\[
t =
\frac{[vw \star 0^{\omega}]_r}{r^{|w|}} + \frac{[u]_r}{r^{|w|}(r^{lk} - 1)}.
\]
Since $1/r^{|w|} \in T_{S''}$ and $1/(r^{lk} - 1) \in T_{S''}$, the
closure and product-stability properties of $T_{S''}$ imply $t \in
T_{S''}$.  \qed

As a consequence, we either have $S'' = \emptyset$ or 
$S'' = \reals_{>0}$, which contradicts our initial assumption
that this set has infinitely many boundary points. As a consequence,
our original set $S$ is definable in $\langle \reals, \ints, +, <\rangle$,
and we have proven Theorem~\ref{theo-cobham-reals-general}.

\section{Conclusions}

In this article, we have established that the sets of real numbers that
can be recognized by finite automata in two sufficiently different
bases are exactly those that are definable in the first-order additive
theory of real and integer variables $\langle \reals, \ints, +,
<\rangle$. In the case of weak deterministic automata, used in actual
implementations of symbolic representation systems~\cite{LASH,FAST,LIRA}, 
the condition on
the bases turns out to be multiplicative independence. It is worth
mentioning that recognizability in multiplicatively dependent bases is
equivalent to recognizability in one of them, and that definability in
$\langle \reals, \ints, +, <\rangle$ implies recognizability in every
base. We have thus obtained a complete characterization of the sets of
numbers recognizable in multiple bases, similar to the one known for
the integer domain~\cite{Cob69}.

For Muller, deterministic B\"uchi, and co-B\"uchi automata,
we have demonstrated that multiplicative
independence of the bases is not a strong enough condition, and that
the bases must have different sets of prime factors in order to force
definability of the represented sets in $\langle \reals, \ints, +,
<\rangle$. Recall that the sets definable in that theory can all be
recognized by weak deterministic automata. We have thus established
that the sets of real numbers that can be recognized by infinite-word 
automata in all encoding bases are exactly those that are
recognizable by weak deterministic automata.

It is worth mentioning that, prior to this result, weak deterministic
automata were already been used as actual data structures for
representing sets of real numbers in state-space exploration
tools~\cite{BJW05,LASH}.  The motivation behind their use was at this
time essentially practical: The algorithmic manipulation of these
automata was considerably simpler than that of unrestricted
infinite-word ones. Moreover, their expressive power was known to be
sufficient for handling the sets definable in $\langle \reals, \ints,
+, <\rangle$, which matched the application requirements. The results
developed in this article now bring an additional theoretical
justification to the choice of weak deterministic automata for
representing sets of real and integer numbers: If recognizability by
automata has to be achieved regardless of the representation base,
then the representable sets are exactly those that can be recognized
by weak deterministic automata.

\bibliography{icalp08} 
\bibliographystyle{alpha}

\end{document}